\documentclass[conference]{IEEEtran}
\IEEEoverridecommandlockouts

\usepackage{cite}
\usepackage{amsmath,amssymb,amsfonts}
\usepackage{graphicx}
\usepackage{textcomp}
\usepackage{xcolor}
\usepackage{bm}
\usepackage{booktabs}
\usepackage{url}
\usepackage{algorithm}
\usepackage{algpseudocode}

\def\BibTeX{{\rm B\kern-.05em{\sc i\kern-.025em b}\kern-.08em
    T\kern-.1667em\lower.7ex\hbox{E}\kern-.125emX}}

\begin{document}

\title{Inter-Area Oscillation Damping in
Data-Center-Integrated Power Systems%
\thanks{This work was partially supported by the Higher Committee
for Education Development in Iraq (HCED). \\
*Corresponding author.}
}

\author{
\IEEEauthorblockN{Ahmed Alfatlawi}
\IEEEauthorblockA{\textit{Electrical and Computer Engineering}\\
\textit{Wayne State University, Detroit, USA}\\
ahmed.alfatlawi@wayne.edu}
\and
\IEEEauthorblockN{Masoud H. Nazari*}
\IEEEauthorblockA{\textit{Electrical and Computer Engineering} \\
\textit{Wayne State University, Detroit, USA}  \\ 
\textit{Electrical and Computer Engineering} \\
\textit{University of Waterloo, Canada} \\ 
masoud.nazari@wayne.edu
}
}

\maketitle

\begin{abstract}
This paper develops explicit dynamic models of a hyperscale data
center, including its heating, ventilation, and air conditioning (HVAC) and uninterruptible power supply (UPS) subsystems, and integrates them
into a small-signal stability framework to investigate the impact
of data center demand response on power system inter-area
oscillations. Through eigenvalue analysis and time-domain
simulations, the results demonstrate that UPS-based demand response
can enhance inter-area oscillation damping. In
contrast, the HVAC subsystem is shown to be inherently incapable
of providing effective oscillation damping due to its limited
thermal response bandwidth. A gradient-based optimization algorithm
is used to tune the UPS controller gain to maximize the damping
ratio of the critical inter-area mode. The
effectiveness of the proposed approach is validated using the
IEEE 39-bus test system.
\end{abstract}


\section{Introduction}

Hyperscale data centers are rapidly transitioning from minor
distribution-level loads into major transmission-level players.
Global data center electricity consumption exceeded 200~TWh in 2022
and is projected to more than double by the end of the
decade~\cite{masanet2020recalibrating,iea2024dc}, driven by cloud
computing, artificial intelligence workloads, and the deployment of
large language models. Individual hyperscale campuses now routinely
demand 100--1{,}000~MW of continuous power, a scale comparable to a
small industrial city.

Unlike conventional industrial loads, which are largely static
or slowly varying, data centers are cyber-physical systems.  Their
internal subsystems (IT server farms, HVAC cooling plants, and
UPS power-conditioning units) exhibit distinct dynamics that
couple back into the grid.  The IT load responds to workload
migrations on a timescale of seconds to minutes; the HVAC
plant tracks room-temperature deviations over several seconds; and
the UPS units can modulate its output within
tens of milliseconds~\cite{sun2022part1,sun2022part2}.  This
hierarchy of timescales creates both a challenge and an
opportunity: while the fast UPS layer is physically capable of
tracking oscillatory grid signals, the slower HVAC layer is not.
The distinction between these two channels is a central theme of
this paper.

Inter-area oscillations are low-frequency (0.1--2~Hz)
electromechanical modes in which groups of generators in different
regions of a transmission network swing against each
other~\cite{kundur1994power,rogers2000power}~\cite{Nazari23}.
These modes become problematic when their damping ratio falls
below approximately 3\%, at which point oscillations decay slowly
and can persist for tens of seconds after a disturbance~\cite{machowski2020power}.  Power system stabilizers
(PSS) have traditionally been the primary tool for restoring
inter-area damping, but PSS installation requires generator-side
hardware upgrades and careful tuning~\cite{sauer2017power, Nazari2012}.  With
the pace of renewable integration accelerating, demand-side
alternatives that can be deployed through software changes are
increasingly attractive.

Prior work on data center demand response has focused primarily
on energy cost minimization~\cite{liu2013data}, peak load
management, and primary frequency regulation~\cite{molina2011demand}.
More recently, researchers have begun to examine the potential of
data centers for reactive power support~\cite{takci2025flexibility},
and small-signal stability improvement~\cite{gyang2025dynamic}.
However, existing stability studies treat the data center as a
single aggregated load without distinguishing between its internal
subsystems.  This approach misses a physical reality:
the HVAC system is unsuitable for inter-area oscillation damping
due to its slow thermal dynamics, while the UPS is well suited
for this purpose. This paper addresses the gaps above with the following
contributions:
\begin{itemize}

    \item Development of detailed dynamic models for both the HVAC
    and UPS subsystems of a hyperscale data center, integrated into
    a comprehensive small-signal stability framework.

    \item Demonstration that UPS-based demand response can enhance inter-area oscillation damping, whereas the HVAC subsystem is inherently ineffective for this purpose
    due to its limited thermal response bandwidth.

    \item Application of a gradient-based optimization approach to
    determine the optimal UPS controller gain that maximizes
    inter-area damping improvement in the IEEE 39-bus benchmark
    system.

\end{itemize}

The remainder of the paper is organized as follows.
Section~II describes the power system configuration with integrated
hyperscale data center loads.  Section~III presents the power system
and data center models, along with the design of the data center
demand response controller.
Section~IV presents the simulation results.  Section~V concludes the overll findings.

\section{Data Center-Integrated Power System Configuration}

\begin{figure}[!t]
\centering
\includegraphics[width=\columnwidth]{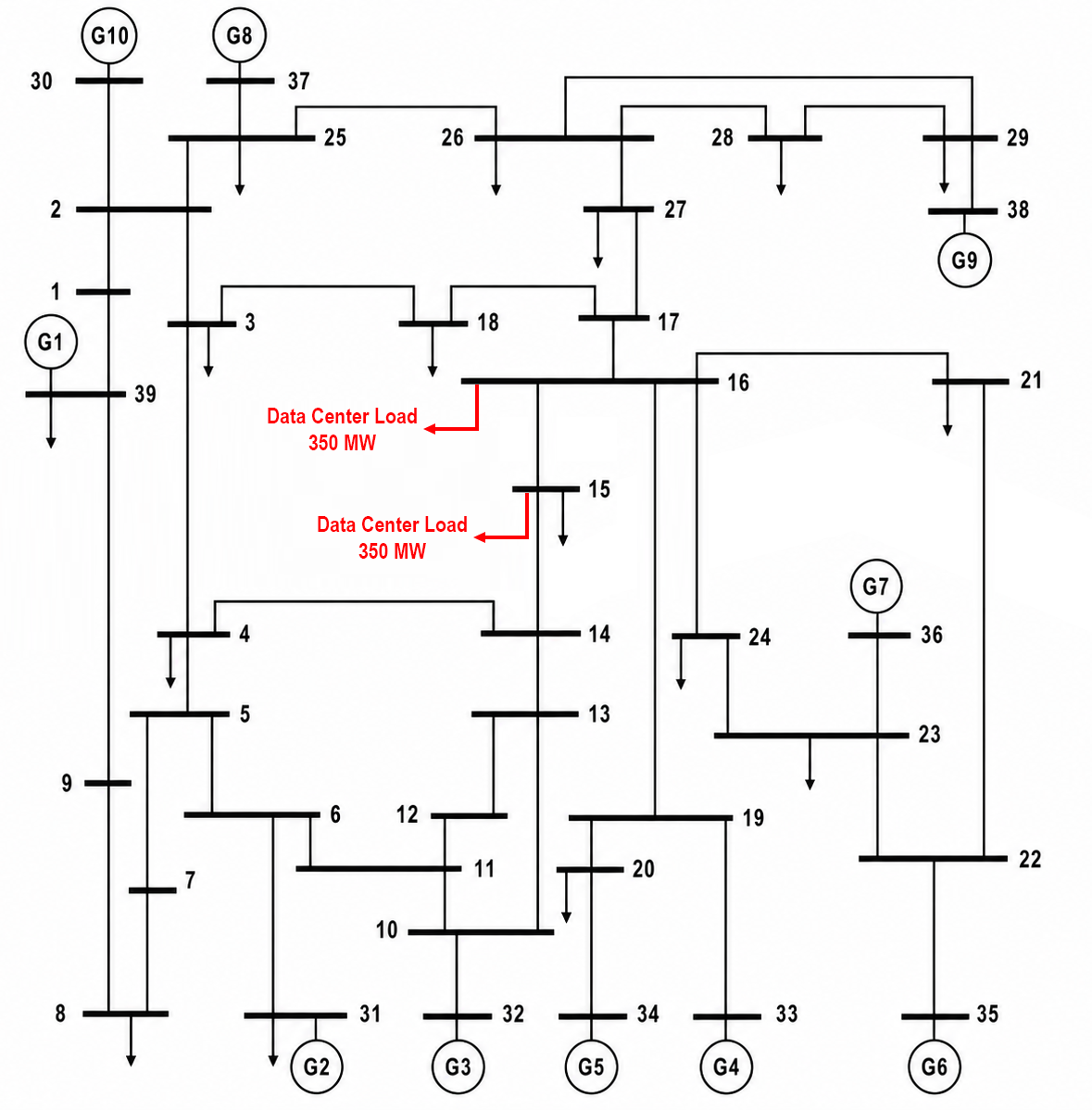}
\caption{IEEE 39-bus New England test system with 700~MW data center
split equally between buses~15 and~16 (350~MW each, shown in red).}
\label{fig:sld}
\end{figure}

The IEEE 39-bus New England test system, shown in
Fig.~\ref{fig:sld}, is used in this paper for inter-area
oscillation studies.  It consists of 10 synchronous generators
connected through a 39-bus, 46-branch transmission network
operating at 345~kV/100~MVA base.  Total system generation
capacity is approximately 6{,}254~MW under base loading
conditions~\cite{rogers2000power,milano2010power}.

For this study, generator damping is set uniformly to $D=2$~pu
across all machines.  The unloaded system exhibits a dominant
inter-area mode at 1.94~Hz with damping ratio $\zeta=1.29\%$.
This mode is selected as the target for the demand-response
controller.

A 700~MW hyperscale data center is connected in equal halves
(350~MW each) at buses~15 and~16.  The total load decomposes as:
\vspace{-0.4em}
\begin{equation}
P_{\mathrm{DC}} = P_{\mathrm{IT}} + P_{\mathrm{HVAC}}
               + P_{\mathrm{aux}} + P_{\mathrm{UPS}} = 700~\mathrm{MW}
\end{equation}
where the IT load $P_{\mathrm{IT}}=420$~MW (60\%), the HVAC load
$P_{\mathrm{HVAC}}=210$~MW (30\%), the auxiliary load
$P_{\mathrm{aux}}=42$~MW (6\%), and the UPS subsystem
$P_{\mathrm{UPS}}=28$~MW (4\%).  These ratios are consistent with
the power usage effectiveness (PUE) values of modern hyperscale
facilities~\cite{ashrae2021tc,patterson2008effect}.

\begin{figure}[!t]
\centering
\includegraphics[width=\columnwidth]{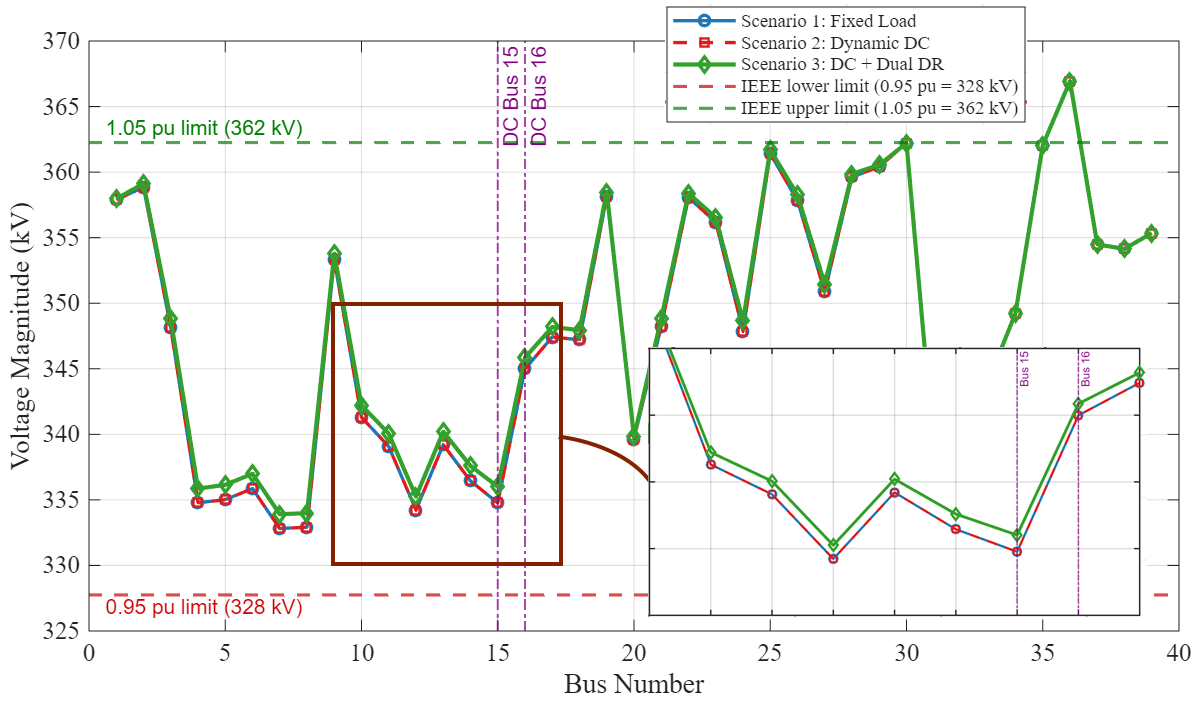}
\caption{Steady-state bus voltage profiles (kV) for all three
scenarios, with zoom inset covering buses~9--17.}
\label{fig:voltage}
\vspace{-0.5em}
\end{figure}

Adding the data center increases the total system load by 11.2\%.  The minimum bus voltage decreases from 0.982~pu
to 0.965~pu at Bus~7, but remains within the 0.95~pu IEEE lower
limit.  Fig.~\ref{fig:voltage} shows the steady-state voltage profile
across all 39 buses.
Scenarios~1 and~2 overlap because they share the same 700~MW
power-flow solution; Scenario~3 shows a small voltage recovery
from demand response reducing effective load by 42~MW.

\vspace{-0.3em}
\subsection{Demand-Response Capability}

Of the four data-center subsystems, only the HVAC and UPS are
considered as potential demand-response resources.  The IT load
is not modulated in this study, as workload throttling introduces
latency and quality-of-service penalties.  The auxiliary load covers lighting, power distribution,
and other infrastructure whose modulation is impractical.

The UPS fast demand-response channel provides $5.6$~MW (20\%
of 28~MW hardware capacity) with a power-electronics time constant
of $\tau_{\mathrm{UPS}}=0.05$~s (50~ms).  The HVAC cooling channel
theoretically offers $42$~MW ($20\%$ of 210~MW), but its
thermal time constant of $T_{\mathrm{HVAC}}=8$~s makes it too slow
for inter-area oscillation damping, as shown analytically in
Section~\ref{sec:controller}.

Three scenarios are evaluated to isolate the effects
of the data center load and the demand-response controller:
Scenario~1 uses a fixed 700~MW static load at buses~15 and~16;
Scenario~2 connects a dynamic 700~MW data center at buses~15 and~16
with demand response disabled; and Scenario~3 activates UPS-based
fast-channel demand response.

\section{Small-Signal Dynamic Modeling and Control Design}
\subsection{Power System Model}

The small-signal stability framework is constructed to capture
interactions between the generators and the
data-center.
Each synchronous generator is represented by a reduced-order model retaining the states most relevant to
inter-area dynamics \cite{Nazari13}.  The rotor-angle and speed deviations
satisfy the classical swing equations:
\begin{align}
\Delta\dot{\delta}_i &= \omega_b\,\Delta\omega_i \\
M_i\,\Delta\dot{\omega}_i &= \Delta P_{m,i} - \Delta P_{e,i}
                             - D_i\,\Delta\omega_i
\end{align}
where $M_i=2H_i/\omega_b$ is the inertia coefficient,
$\omega_b=2\pi\times60$~rad/s is the base angular frequency,
$\Delta P_{m,i}$ is the mechanical power deviation, and
$\Delta P_{e,i}$ is the electrical power deviation.
Governor dynamics relate mechanical power to speed deviation:
\begin{equation}
T_{g,i}\,\Delta\dot{P}_{m,i} = -\Delta P_{m,i}
   - \frac{1}{R_i}\Delta\omega_i + \Delta P_{\mathrm{ref},i}
\end{equation}
and excitation-system dynamics track terminal voltage:
\begin{equation}
T_{A,i}\,\Delta\dot{E}_{fd,i} = -\Delta E_{fd,i}
   + K_{A,i}\bigl(\Delta V_{\mathrm{ref},i} - \Delta V_i\bigr)
\end{equation}
The 10-machine model yields 40 state variables,
with inertia constants $H_i\in[3.45,5.00]$~s drawn from the
standard IEEE 39-bus dataset~\cite{rogers2000power} and
$D_i=2$~pu for all generators.

The transmission network couples generators through the
admittance matrix $Y_{\mathrm{bus}}$.  Active and reactive power
injections at bus $i$ are related to voltages at buses by:
\begin{align}
P_i &= V_i\sum_{j} V_j\bigl(G_{ij}\cos\theta_{ij}
        + B_{ij}\sin\theta_{ij}\bigr) \\
Q_i &= V_i\sum_{j} V_j\bigl(G_{ij}\sin\theta_{ij}
        - B_{ij}\cos\theta_{ij}\bigr)
\end{align}
where $G_{ij}+jB_{ij}$ are the elements of $Y_{\mathrm{bus}}$.
Linearizing these expressions at the power-flow operating point
yields the network Jacobian matrices used in the generator and
load coupling blocks.

The electrical proximity of each generator to the data-center
buses is quantified by electrical-distance coupling weights
derived from the bus impedance matrix
$Z_{\mathrm{bus}}=Y_{\mathrm{bus}}^{-1}$:
\begin{equation}
w_i = \frac{1/d_i}{\sum_{k=1}^{N_g} 1/d_k}, \quad d_i = |Z_{i,\mathrm{dc}}|
\end{equation}
where $|Z_{i,\mathrm{dc}}|$ is the magnitude of the transfer
impedance between generator bus $i$ and the data-center buses. The electrical proximity information can be used for eigenvalue sensitivity analysis.

\subsection{Data Center Dynamic Model}
\label{sec:dcmodel}

The three-timescale structure of the data center model is a key
contribution of this work.  Each subsystem operates at a
 different timescale: IT load dynamics span 1--10~s,
HVAC thermal dynamics span 5--30~s, and UPS power-electronics
dynamics span 10--100~ms.  This separation of timescales means
that the subsystems respond to different frequency ranges of grid
disturbances and must be modeled independently. Table~\ref{tab:params} summarises
all model parameters.

\begin{table}[!t]
\renewcommand{\arraystretch}{1.2}
\caption{Data Center Dynamic Model Parameters}
\label{tab:params}
\centering
\begin{tabular}{llll}
\toprule
\textbf{Parameter} & \textbf{Symbol} & \textbf{Value} & \textbf{Unit} \\
\midrule
\multicolumn{4}{l}{\textit{IT Load}} \\
IT time constant      & $T_{\mathrm{IT}}$   & 2.0  & s \\
Freq.\ sensitivity    & $K_{\omega,\mathrm{IT}}$ & 0.0 & --- \\
\multicolumn{4}{l}{\textit{HVAC Model}} \\
HVAC time constant    & $T_{\mathrm{HVAC}}$ & 8.0  & s \\
Thermal capacitance   & $C_{\mathrm{th}}$   & 5.0  & pu \\
Thermal resistance    & $R_{\mathrm{th}}$   & 1.0  & pu \\
Thermostatic gain     & $K_{\mathrm{th}}$   & 1.8  & --- \\
HVAC COP              & $\gamma$            & 2.0  & --- \\
DR range              & $\Delta P_{\mathrm{HVAC}}$ & $\pm$42 & MW \\
\multicolumn{4}{l}{\textit{UPS Model}} \\
UPS time constant     & $\tau_{\mathrm{UPS}}$ & 0.05 & s \\
Lead compensator      & $T_{\mathrm{lead}}$ & 0.085 & s \\
Lag compensator       & $T_{\mathrm{lag}}$  & 0.020 & s \\
UPS DR range          & $\Delta P_{\mathrm{UPS}}$ & $\pm$5.6 & MW \\
Selected gain         & $K_{\mathrm{UPS}}$  & 3571 & --- \\
\multicolumn{4}{l}{\textit{System Base}} \\
Base power            & $S_{\mathrm{base}}$ & 100  & MVA \\
Base frequency        & $f_{\mathrm{base}}$ & 60   & Hz \\
\bottomrule
\end{tabular}
\end{table}

The IT load represents the computational power consumption of the
server infrastructure and is the dominant subsystem.  In steady
state, IT power varies with workload utilization $r\in[0,1]$
through a nonlinear relationship~\cite{liu2013data,ecocenter2026}:
\begin{equation}
P_{\mathrm{IT}} = P_{\mathrm{idle}} + \alpha r + \beta r^2
\label{eq:IT_ss}
\end{equation}
where $P_{\mathrm{idle}}$ is the no-load power and $\alpha$, $\beta$ are empirically
determined utilization coefficients.  The dynamic model captures
the response of IT power to changes in grid frequency, terminal
voltage, and workload:
\begin{equation}
T_{\mathrm{IT}}\,\Delta\dot{P}_{\mathrm{IT}}
= -\Delta P_{\mathrm{IT}} + K_{\mathrm{IT}}\Delta r
  + K_{v,\mathrm{IT}}\Delta V + K_{\omega,\mathrm{IT}}\Delta\omega
\label{eq:IT_dyn}
\end{equation}
The time constant $T_{\mathrm{IT}}$ reflects the thermal
inertia of the server and the response latency of power
management firmware.  In this study, the IT load is retained as
a non-controllable dynamic load.

The HVAC subsystem represents the primary thermal
mass, consisting of computer room air handlers, chillers,
and cooling towers.
The HVAC cooling-power dynamics are modeled as a first-order
thermofluid process driven by room-temperature deviation and the
demand-response control signal~\cite{ashrae2021tc,patterson2008effect}:
\begin{equation}
T_{\mathrm{HVAC}}\,\Delta\dot{P}_{\mathrm{HVAC}}
  = -\Delta P_{\mathrm{HVAC}}
     + K_{\mathrm{th}}\,\Delta T_{\mathrm{room}}
     + \Delta P_{\mathrm{DR,slow}}
\label{eq:hvac}
\end{equation}
The room-temperature deviation evolves from the thermal energy
balance of the server hall, balancing IT heat generation against
HVAC heat removal and natural heat dissipation:
\begin{equation}
C_{\mathrm{th}}\,\Delta\dot{T}_{\mathrm{room}}
  = \Delta P_{\mathrm{IT}}
     - \frac{1}{R_{\mathrm{th}}}\Delta T_{\mathrm{room}}
     - \gamma\,\Delta P_{\mathrm{HVAC}}
\label{eq:troom}
\end{equation}
Equations~\eqref{eq:hvac} and~\eqref{eq:troom} together form a
coupled two-state thermal subsystem ($\Delta P_{\mathrm{HVAC}}$ and
$\Delta T_{\mathrm{room}}$) and two exogenous inputs
($\Delta P_{\mathrm{IT}}$ and $\Delta P_{\mathrm{DR,slow}}$).
The allowable HVAC demand-response range is $20\%$ of
nominal power~\cite{ashrae2021tc}.

The UPS subsystem is the smallest power consumer
but the most capable of rapid response.  Modern data-center UPS
systems use double-conversion topology in which a PWM-controlled
inverter can modulate output power within tens of milliseconds.
The UPS actuator is modeled as a first-order system:
\begin{equation}
\tau_{\mathrm{UPS}}\,\Delta\dot{P}_{\mathrm{UPS}}
  = -\Delta P_{\mathrm{UPS}} + \Delta P_{\mathrm{DR,fast}}
\label{eq:ups}
\end{equation}
where $\Delta P_{\mathrm{DR,fast}}$ is the demand-response control
signal generated by the Center-of-Inertia (COI)-based controller
described in Section~\ref{sec:controller}.  The UPS demand-response
range is 20\% of hardware capacity.  This headroom
of 80\% is preserved for the primary UPS mission of
providing ride-through power during grid faults or momentary
voltage sags.

The complete small-signal model combines all subsystem states into
a 47-state augmented system:
\begin{equation}
\Delta x =
\bigl[\Delta x_g^T\ \Delta x_{\mathrm{dc}}^T\
\Delta x_{\mathrm{ctrl}}^T\ \Delta P_{\mathrm{UPS}}\bigr]^T
\in \mathbb{R}^{47}
\end{equation}
where $\Delta x_g\in\mathbb{R}^{40}$ contains the 10-generator
states; $\Delta x_{\mathrm{dc}}\in\mathbb{R}^3$
contains the data-center states
($\Delta P_{\mathrm{IT}}$,
$\Delta P_{\mathrm{HVAC}}$, and
$\Delta T_{\mathrm{room}}$);
$\Delta x_{\mathrm{ctrl}}\in\mathbb{R}^3$ contains the slow
controller states; and $\Delta P_{\mathrm{UPS}}\in\mathbb{R}^1$
is the UPS actuator state.  The governing equation is
$\Delta\dot{x}=A_{\mathrm{sys}}\Delta x$, where the augmented
system matrix has the block structure:
\begin{equation}
A_{\mathrm{sys}} =
\begin{bmatrix}
A_g & A_{g,\mathrm{dc}} & 0 \\
A_{\mathrm{dc},g} & A_{\mathrm{dc}} & A_{\mathrm{dc},c} \\
A_{c,g} & 0 & A_c
\end{bmatrix}
\end{equation}
The off-diagonal blocks $A_{g,\mathrm{dc}}$ and $A_{g,\mathrm{dc}}$
encode the bidirectional electrical coupling between the generators
and the data-center through the network admittance matrix.

\subsection{UPS Demand-Response Controller}
\label{sec:controller}

The starting point for the controller design is a formal derivation
of the damping contribution that a frequency-responsive
demand-response channel makes to the swing equation.  Consider
the aggregated swing equation for the dominant inter-area mode,
where $\Delta P_{\mathrm{DR}}$ represents the data-center
demand-response power injection:
\begin{equation}
M\,\Delta\dot{\omega} = \Delta P_m - \Delta P_e
  - \Delta P_{\mathrm{base}} + \Delta P_{\mathrm{DR}} - D\,\Delta\omega
\label{eq:swing}
\end{equation}
Here $\Delta P_{\mathrm{base}}$ represents the power
variation of the data-center load, $D$ is the natural generator
damping, and $\Delta P_{\mathrm{DR}}$ is the active demand-response
contribution. The demand-response controller generates a signal
proportional to the COI frequency deviation, denoted
$\Delta\omega_{\mathrm{COI}}$:
\begin{equation}
\Delta P_{\mathrm{DR}} = -K_{\mathrm{DR},\omega}\,\Delta\omega_{\mathrm{COI}}
\approx -K_{\mathrm{DR},\omega}\,\Delta\omega
\end{equation}
Substituting into~\eqref{eq:swing} yields:
\begin{equation}
M\,\Delta\dot{\omega} = \Delta P_m - \Delta P_e - \Delta P_{\mathrm{base}}
  - \bigl(D + D_{\mathrm{DR}}\bigr)\,\Delta\omega
\end{equation}
The demand-response contribution appears as an additive virtual
damping coefficient $D_{\mathrm{DR}} = K_{\mathrm{DR},\omega}$.
This result implies that the contribution depends only on the
controller gain.

The UPS transfer function
$G_{\mathrm{UPS}}(s)=1/(1+s\tau_{\mathrm{UPS}})$ introduces a
phase lag of $-31.4^\circ$ at 1.94~Hz, reducing the effective
damping contribution.  A lead-lag compensator with
$T_{\mathrm{lead}}=0.085$~s and $T_{\mathrm{lag}}=0.020$~s
corrects the phase to approximately $0^\circ$, maximizing the
real part of the compensated transfer function at the critical
frequency. 

The HVAC thermal dynamics act as a first-order low-pass filter
between the demand-response command and the actual cooling power
delivered.  At the 1.94~Hz critical inter-area frequency, the
filter magnitude is:
\vspace{-0.4em}
\[
\bigl|G_{\mathrm{HVAC}}\bigr|_{1.94\,\mathrm{Hz}}
= \frac{1}{\sqrt{1+(2\pi \cdot 1.94 \cdot 8.0)^2}} = 0.0103 .
\]
\vspace{-0.2em}
This 98.97\% signal attenuation is a consequence
of the thermal time constant $T_{\mathrm{HVAC}}=8$~s being 15.5
times larger than the inter-area oscillation period
$T_{\mathrm{osc}}=0.515$~s.  The HVAC system therefore cannot
respond to the 1.94~Hz oscillation signal due to
physical bandwidth limitation.  This is confirmed by the
simulation results in Section~\ref{sec:results}.

\subsection{Gradient-Based UPS Gain Optimization}
\label{sec:optimal_gain}


To maximize the minimum damping ratio of the critical inter-area oscillation mode, the UPS controller gain $K_{\mathrm{UPS}}$ is optimally determined by solving the following problem:

\begin{equation}
\underset{K_{\mathrm{UPS}}}{\text{maximize}} \quad
\zeta_{\min}\!\bigl(A_{\mathrm{sys}}(K_{\mathrm{UPS}})\bigr)
\label{eq:opt_problem}
\end{equation}
where $\zeta_{\min}(A_{\mathrm{sys}}) =
\min_i \bigl[-\mathrm{Re}(\lambda_i)/|\lambda_i|\bigr]$
is the minimum damping ratio over all eigenvalues
$\lambda_i$ of $A_{\mathrm{sys}}$. Problem~\eqref{eq:opt_problem} can
be written as the following minimization:
\begin{equation}
\underset{K_{\mathrm{UPS}}}{\text{minimize}} \quad
J(K_{\mathrm{UPS}}) = -\zeta_{\min}\!\bigl(A_{\mathrm{sys}}(K_{\mathrm{UPS}})\bigr)
\label{eq:opt_min}
\end{equation}

Subject to the following constraints:

\begin{equation}
K_{\min} \leq K_{\mathrm{UPS}} \leq K_{\max}
\end{equation}

\begin{equation}
|\Delta P_{\mathrm{UPS}}| \leq 0.8P_{\mathrm{UPS}}
\label{eq:ups_limit}
\end{equation}

\begin{equation}
\mathrm{Re}(\lambda_i) < 0, \quad \forall\, i = 1,\ldots,47
\end{equation}

\begin{equation}
\label{cons4}
\mathrm{Re}\!\bigl(G_{\mathrm{UPS,comp}}(j\omega_{\mathrm{crit}})\bigr) > 0
\end{equation}
Constraint \eqref{cons4} ensures that the
demand-response injection contributes positive
virtual damping.
The gradient of the objective function with respect to the
decision variable is computed using the eigenvalue
sensitivity formula~\cite{machowski2020power}:
\begin{equation}
\frac{\mathrm{d}J}{\mathrm{d}K_{\mathrm{UPS}}}
= -\frac{\mathrm{d}\zeta_{\min}}{\mathrm{d}K_{\mathrm{UPS}}}
= -\mathrm{Re}\!\left(
    \frac{\phi_c^T \,
          \dfrac{\partial A_{\mathrm{sys}}}{\partial K_{\mathrm{UPS}}}
          \, \psi_c}
         {\phi_c^T \psi_c}
  \right) \cdot
  \frac{\partial \zeta}{\partial \lambda_c}
\label{eq:gradient}
\end{equation}
where $\lambda_c$ is the critical inter-area eigenvalue,
$\psi_c$ and $\phi_c$ are its right and left eigenvectors,
and $\partial A_{\mathrm{sys}}/\partial K_{\mathrm{UPS}}$ is the
sparse matrix of partial derivatives.
The gradient-based gain optimization procedure is
presented as Algorithm~1, and the optimization results are summarized in 
Table~\ref{tab:search}.  %
\begin{algorithm}[!t]
\caption{Gradient-Based UPS Gain Optimization}
\label{alg:grad_opt}
\small
\begin{algorithmic}[1]
\State \textbf{Initialize:} $K \leftarrow K_0$,
       $\zeta^* \leftarrow 0$, $K^* \leftarrow K_0$,
       step size $\eta$, tolerance $\epsilon$
\Repeat
    \State Build $A_{\mathrm{sys}}(K)$ (47-state augmented matrix)
    \State Compute eigenvalues $\{\lambda_i\}_{i=1}^{47}$
           of $A_{\mathrm{sys}}(K)$
    \State \textbf{Check stability:} if $\exists\,\mathrm{Re}(\lambda_i)\geq0$,
           reduce $K$ and go to Step 3
    \State Identify critical inter-area eigenvalue
           $\lambda_c = \arg\min_i \zeta_i$
           (for $f_i \in [0.1,\,2]$~Hz)
    \State Compute $\zeta_{\min} = -\mathrm{Re}(\lambda_c)/|\lambda_c|$
    \State Compute gradient
           $g \leftarrow
            \mathrm{d}\zeta_{\min}/\mathrm{d}K$
           using \eqref{eq:gradient}
    \State Update gain: $K \leftarrow K + \eta\,g$
           \hfill (gradient ascent)
    \State Enforce constraints:
           $K \leftarrow \min\!\bigl(\max(K,\,K_{\min}),\,K_{\max}\bigr)$
    \If{$\zeta_{\min} > \zeta^*$}
        \State $\zeta^* \leftarrow \zeta_{\min}$;
               $K^* \leftarrow K$
    \EndIf
    \State Adapt step size $\eta$ (Armijo line search)
\Until{$|\Delta\zeta_{\min}| < \epsilon$}
\State \textbf{Return:} $K_{\mathrm{UPS}} \leftarrow K^*$
\end{algorithmic}
\end{algorithm}


\begin{table}[!t]
\renewcommand{\arraystretch}{1.2}
\setlength{\tabcolsep}{2pt}
\caption{Optimization-Based Gain Selection: Representative
Near-Optimal Operating Points}
\label{tab:search}
\centering
\footnotesize
\begin{tabular}{@{}p{4.25cm}ccc@{}}
\toprule
\textbf{Operating Point} & $P_{\mathrm{pu}}{\times}K$
  & $\mathrm{Re}(G)$ & $\zeta_{\min}$ (\%) \\
\midrule
Nominal ($K=3571$, $P=5.6$~MW)     & 200.0 & 1.194 & 2.240 \\
Near-equivalent validation point ($K=375$, $P=50$~MW) & 187.5 & 1.194 & 2.243 \\
\bottomrule
\end{tabular}
\end{table}

\section{Simulation Results}
\label{sec:results}

\begin{figure}[!t]
\centering
\includegraphics[width=\columnwidth]{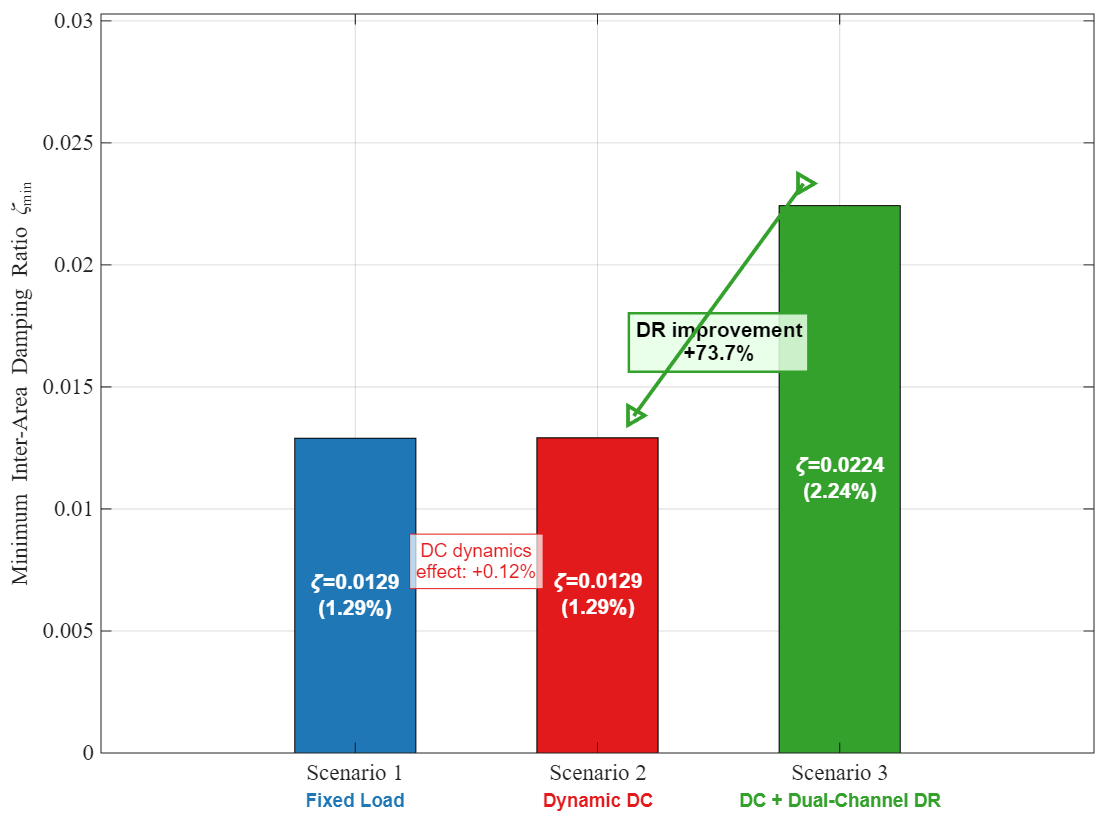}
\caption{Minimum inter-area damping ratio across the three studied
scenarios.}
\label{fig:damping_bar}
\end{figure}

\begin{figure}[!t]
\centering
\includegraphics[width=\columnwidth]{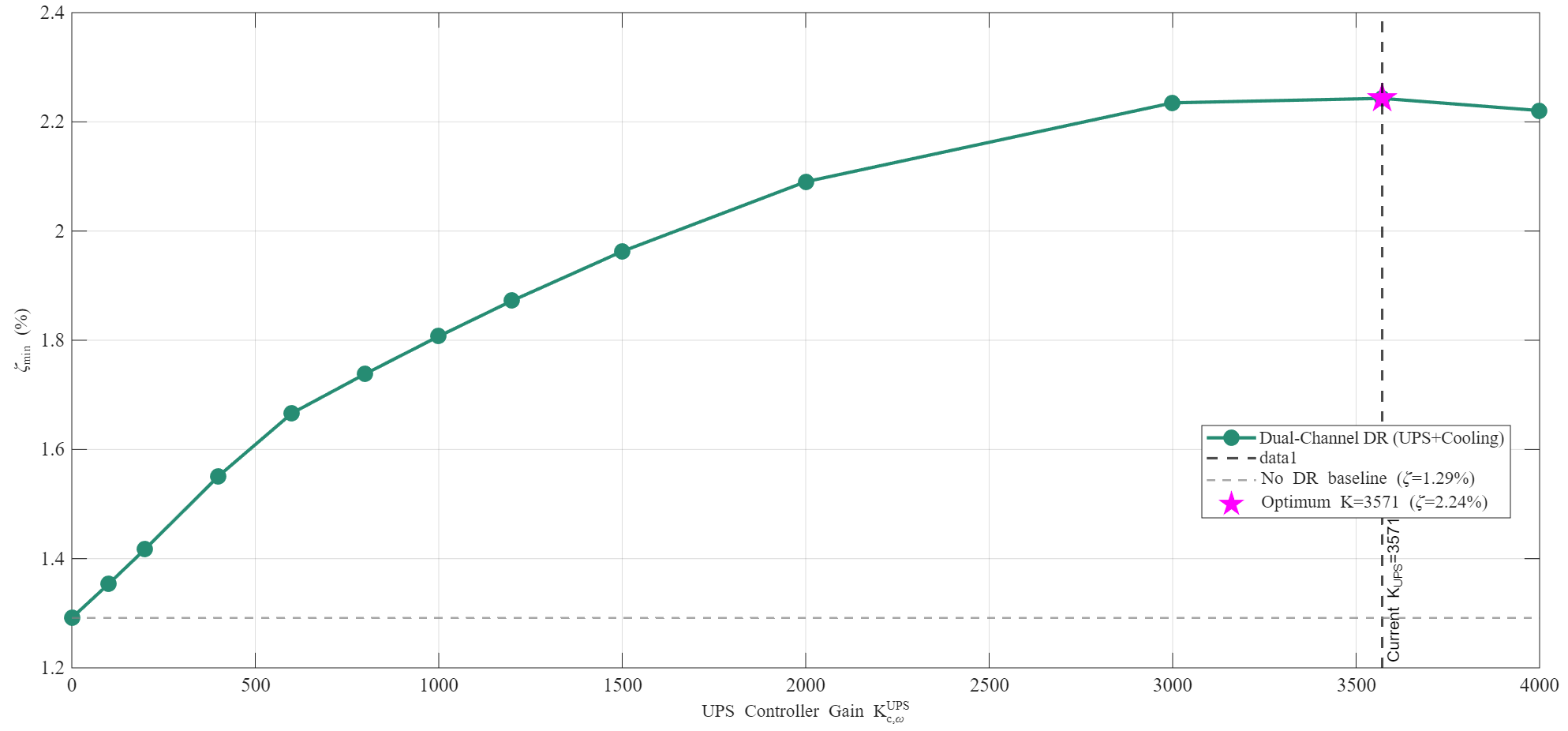}
\caption{Sensitivity of $\zeta_{\min}$ to UPS gain $K_{\mathrm{UPS}}$
at $P=5.6$~MW.}
\label{fig:sensitivity}
\vspace{-0.6em}
\end{figure}

We compare the system damping characteristics under the three
scenarios described in Section~II.
Comparing Scenarios~1 and~2 shows that dynamically connecting
the data center changes $\zeta$ by only 0.12\%---the data center
is power-flow neutral without demand response.  The 73.7\%
improvement from Scenario~2 to~3 is achieved by the additive
virtual damping provided by the UPS demand-response channel.
Fig.~\ref{fig:damping_bar} shows the damping ratio comparison.
%
Fig.~\ref{fig:sensitivity} also shows the sensitivity of $\zeta_{\min}$
to the UPS gain $K$ near the operating point.


For the time-domain analysis, a 100~MW load step disturbance is
applied at buses~15 and~16 at $t = 0$~s. The resulting dynamic
performance is evaluated using several key metrics, which are
summarized in Table~\ref{tab:timedomain}.

\begin{table}[!t]
\renewcommand{\arraystretch}{1.2}
\setlength{\tabcolsep}{2pt}
\caption{Time-Domain Performance Metrics (100~MW Step Disturbance)}
\label{tab:timedomain}
\centering
\footnotesize
\begin{tabular}{@{}lccc@{}}
\toprule
\textbf{Metric} & \textbf{S2 (no DR)} & \textbf{S3 (DR)} & \textbf{Impr.} \\
\midrule
Settling time (s)      & 14.39 & 10.48 & $-$27.2\% \\
Peak COI frequency deviation (Hz)  & 0.2233 & 0.1929 & $-$13.6\% \\
Frequency nadir (Hz)   & 59.777 & 59.807 & +0.030 \\
Modal energy           & $3.678\!\times\!10^{-3}$ & $1.808\!\times\!10^{-3}$ & $-$50.9\% \\
Peak UPS DR (MW)       & ---  & 5.600 & 100\% sat. \\
Peak HVAC DR (MW)      & ---  & 0.000 & 0\% \\
\bottomrule
\end{tabular}
\end{table}

The UPS channel saturates immediately at its $5.6$~MW hardware
limit, confirming full actuator utilization and that the virtual
damping coefficient $D_{\mathrm{DR}}$ operates at its maximum.
The HVAC channel contributes zero demand response
throughout the entire transient, confirming
the bandwidth attenuation demonstrated analytically in
Section~\ref{sec:controller}.

\begin{figure}[!t]
\centering
\includegraphics[width=\columnwidth]{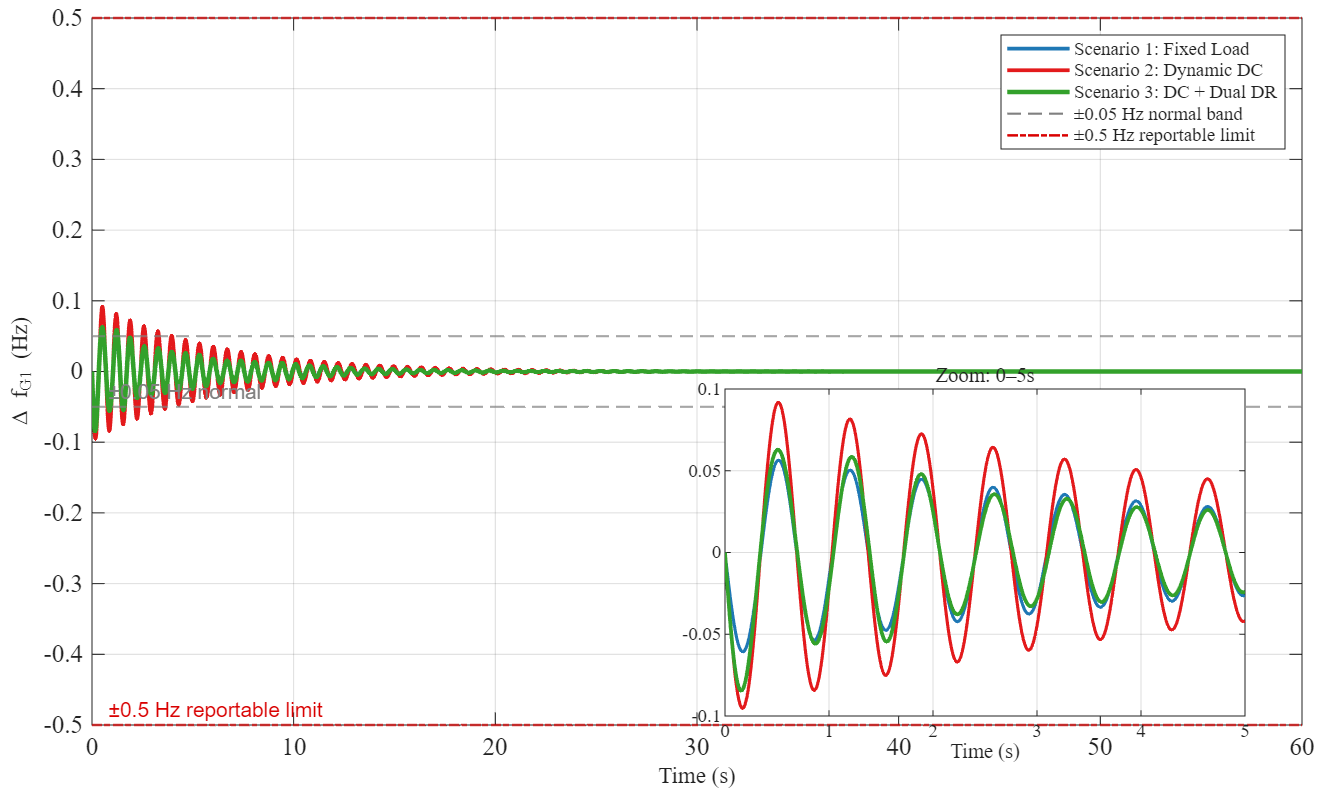}
\caption{Generator~1 frequency deviation following the 100~MW
load-step disturbance.}
\label{fig:gen1freq}
\end{figure}

\begin{figure}[!t]
\centering
\includegraphics[width=\columnwidth]{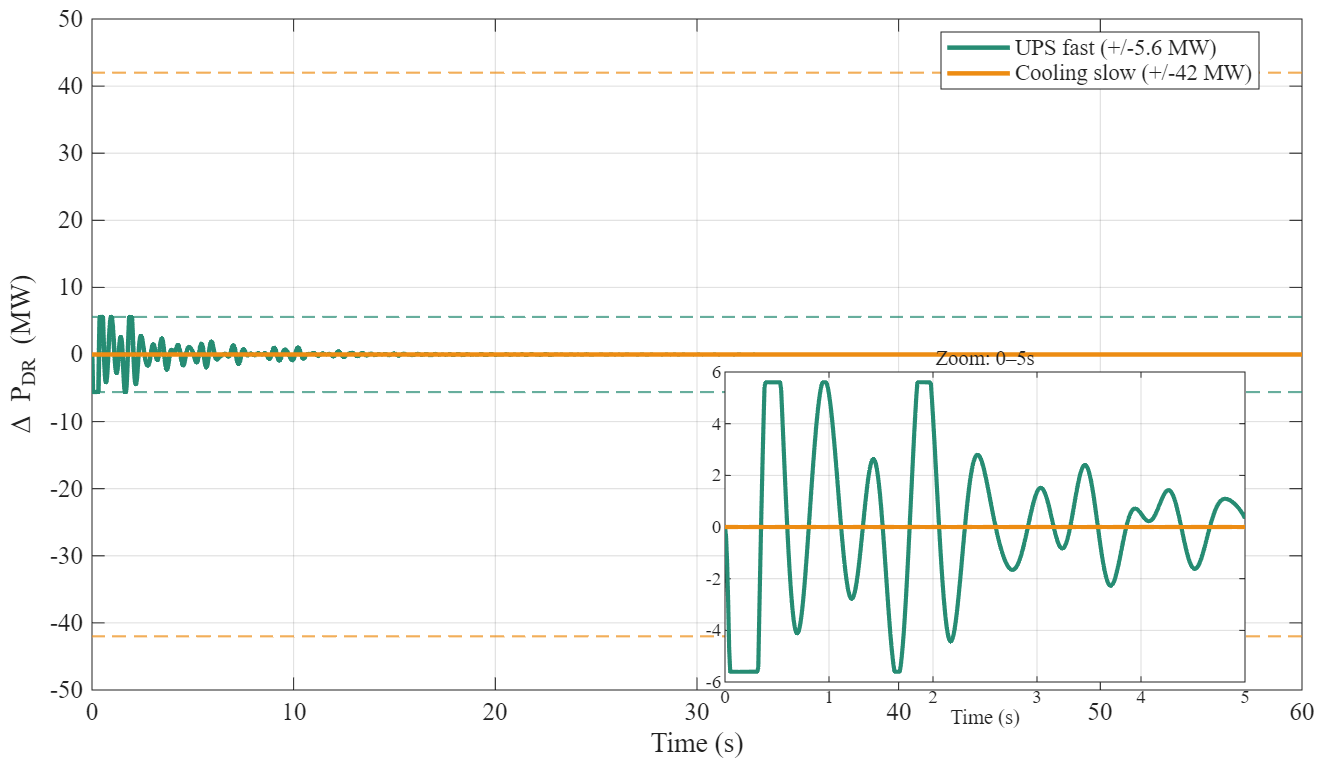}
\caption{UPS demand-response power output during Scenario~3.}
\label{fig:drpower}
\vspace{-0.6em}
\end{figure}

Fig.~\ref{fig:gen1freq} shows the Generator~1 frequency deviation.
The zoom inset on the first 5~s reveals that Scenario~3
(green, with UPS DR) decays faster than Scenario~2 (red, no DR).
Fig.~\ref{fig:drpower} also shows the DR channel outputs explicitly.

\begin{figure}[!t]
\centering
\includegraphics[width=\columnwidth]{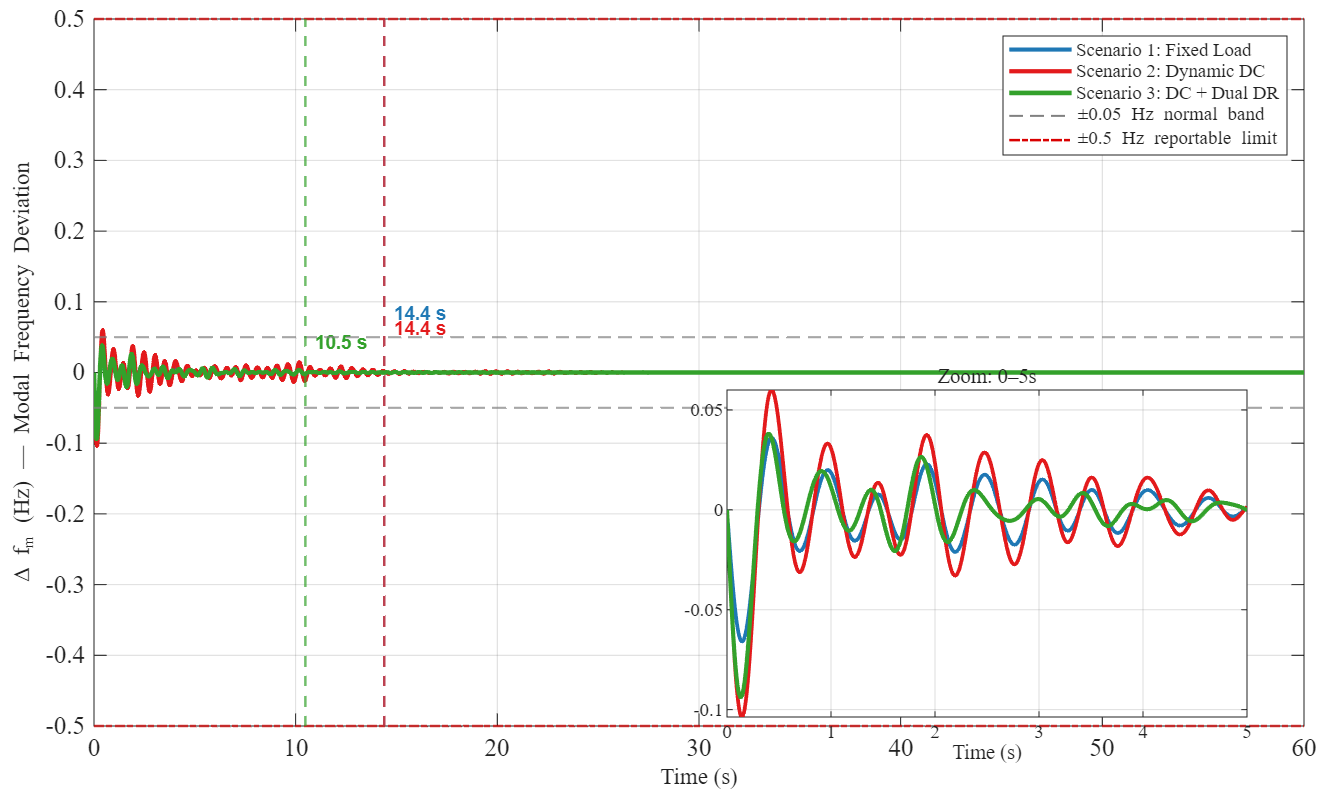}
\caption{The COI frequency response following the
100~MW load-step disturbance.}
\label{fig:modal_freq}
\end{figure}

\begin{figure}[!t]
\centering
\includegraphics[width=\columnwidth]{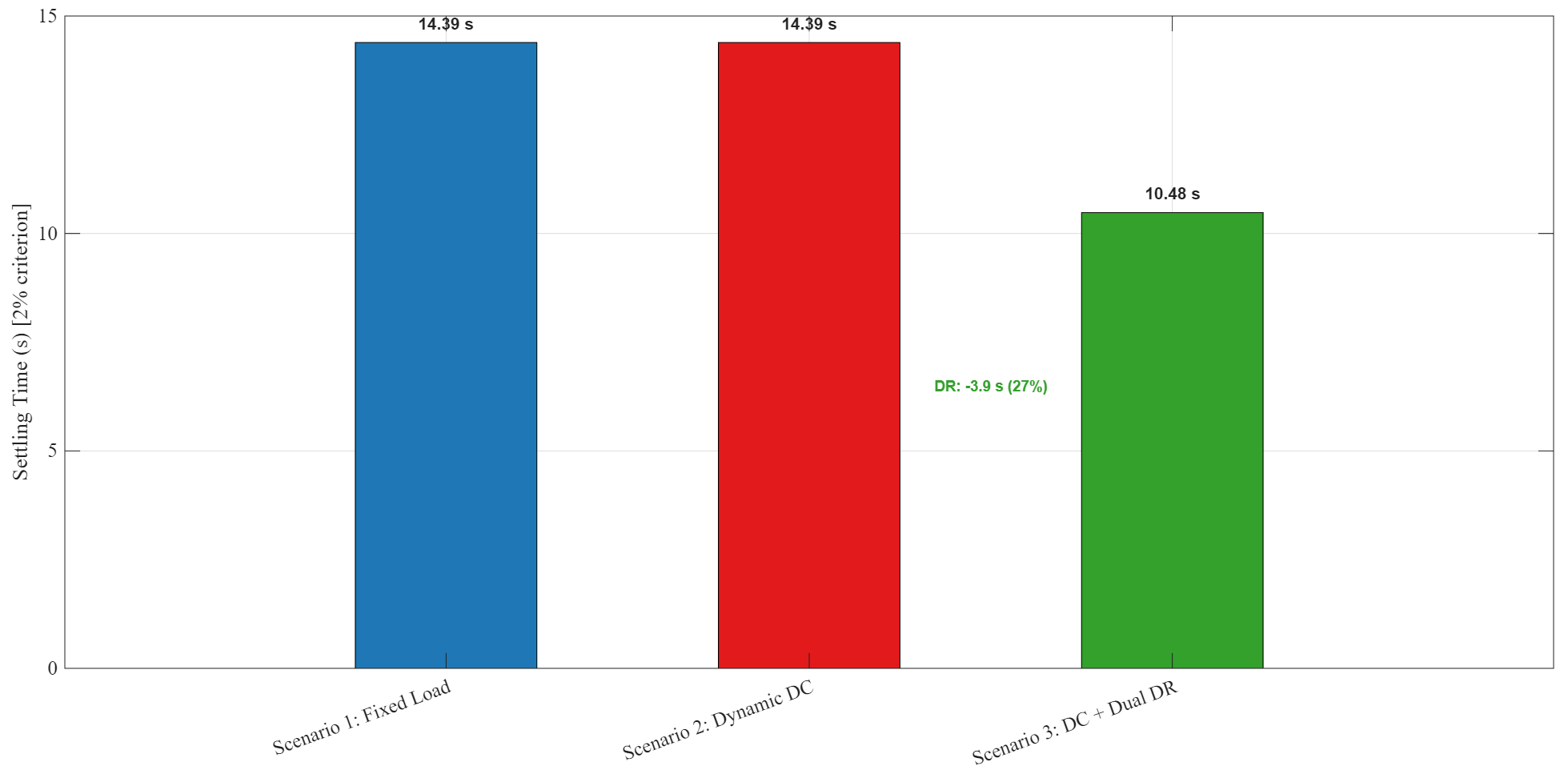}
\caption{Settling-time comparison using the 2\% criterion.}
\label{fig:settling}
\vspace{-0.6em}
\end{figure}

In addition, Fig.~\ref{fig:modal_freq} presents the COI frequency response
following the disturbance. In the absence of demand response, the
first oscillatory excursion exceeds the $\pm0.05$~Hz nominal
frequency-deviation band. But the UPS-based demand response suppresses the oscillations and reduces the settling
time from 14.39~s to 10.48~s. Figure~\ref{fig:settling} further
compares the settling-time performance under different demand
response scenarios.

\vspace{-0.5em}

\section{Conclusion}

This paper presented a small-signal stability analysis of a
data-center-integrated power system with explicit HVAC and UPS
dynamic models.
The analytical results show that
the HVAC subsystem is physically incapable of contributing to
inter-area oscillation damping, due to slow response. 
This is a hardware constraint, not a tuning deficiency.
The UPS subsystem, by contrast, contributes additive virtual
damping to the generator swing equation via its fast
power-electronics response.  At the optimal gain determined by the
gradient-based optimization procedure, the UPS subsystem achieves
a 73.7\% improvement in inter-area damping ratio.
This improvement can be achieved through a simple firmware-level
gain adjustment, requiring no additional hardware investment.

Future work will investigate adding an independent fast
demand-response channel through IT workload throttling, which is
expected to further improve inter-area damping beyond the single-channel
results demonstrated here. Additional studies may also investigate
adaptive controller tuning and sensitivity-aware parameter selection
under varying grid and data-center operating conditions.



\begin{thebibliography}{00}

\bibitem{masanet2020recalibrating}
E.~Masanet, A.~Shehabi, N.~Lei, S.~Smith, and J.~Koomey,
``Recalibrating global data center energy-use estimates,''
\emph{Science}, vol.~367, no.~6481, pp.~984--986, Feb.~2020.

\bibitem{iea2024dc}
International Energy Agency, ``Data Centres and Data Transmission
Networks,'' IEA, Paris, Tech. Rep., 2024. [Online]. Available:
\url{https://www.iea.org/reports/data-centres-and-data-transmission-networks}

\bibitem{sun2022part1}
J.~Sun, M.~Xu, M.~Cespedes, and M.~Kauffman,
``Data center power system stability---Part~I: Power supply impedance
modeling,'' \emph{CSEE J. Power Energy Syst.}, vol.~8, no.~2,
pp.~403--419, Mar.~2022.

\bibitem{sun2022part2}
J.~Sun, M.~Xu, M.~Cespedes, and M.~Kauffman,
``Data center power system stability---Part~II: System modeling and
analysis,'' \emph{CSEE J. Power Energy Syst.}, vol.~8, no.~2,
pp.~420--431, Mar.~2022.

\bibitem{kundur1994power}
P.~Kundur, \emph{Power System Stability and Control}.
New York, NY, USA: McGraw-Hill, 1994.

\bibitem{rogers2000power}
G.~Rogers, \emph{Power System Oscillations}.
Norwell, MA, USA: Kluwer Academic, 2000.

\bibitem{Nazari23}
S.~Xie, M.~H.~Nazari, and L.~Y.~Wang,
``Learning-based distributed optimal power sharing and frequency
control under cyber contingencies,''
\emph{Int. J. Electr. Power Energy Syst.}, vol.~152, 2023.

\bibitem{machowski2020power}
J.~Machowski, Z.~Lubosny, J.~W.~Bialek, and J.~R.~Bumby,
\emph{Power System Dynamics: Stability and Control}, 3rd~ed.
Hoboken, NJ, USA: Wiley, 2020.

\bibitem{sauer2017power}
P.~W.~Sauer, M.~A.~Pai, and J.~H.~Chow,
\emph{Power System Dynamics and Stability}, 2nd~ed.
Hoboken, NJ, USA: Wiley-IEEE Press, 2017.

\bibitem{Nazari2012}
M. H. Nazari, M. Ilic, J. P. Lopes,
Small-signal stability and decentralized control design for electric energy systems with a large penetration of distributed generators,
Control Engineering Practice,
Volume 20, Issue 9,
2012,
Pages 823-831.


\bibitem{liu2013data}
Z.~Liu, I.~Liu, S.~Low, and A.~Wierman,
``Pricing data center demand response,''
in \emph{Proc. ACM SIGMETRICS}, Pittsburgh, PA, USA, 2013,
pp.~111--122.

\bibitem{molina2011demand}
J.~D.~Molina-Garcia, F.~Bouffard, and D.~S.~Kirschen,
``Decentralized demand-side contribution to primary frequency control,''
\emph{IEEE Trans. Power Syst.}, vol.~26, no.~1, pp.~411--419,
Feb.~2011.

\bibitem{takci2025flexibility}
M.~T.~Takci, M.~Qadrdan, J.~Summers, and J.~Gustafsson,
``Data centres as a source of flexibility for power systems,''
\emph{Energy Rep.}, vol.~13, pp.~3661--3671, 2025.

\bibitem{gyang2025dynamic}
P.~P.~Gyang, P.~Chakraborty, L.~Meegahapola, and X.~Yu,
``Dynamic modeling of a data center for power system stability
studies,'' \emph{IEEE Trans. Power Syst.}, 2025.

\bibitem{milano2010power}
F.~Milano, \emph{Power System Modelling and Scripting}.
London, UK: Springer, 2010.

\bibitem{ashrae2021tc}
ASHRAE Technical Committee 9.9,
\emph{Data Center Power Equipment Thermal Guidelines and Best
Practices}. Atlanta, GA, USA: ASHRAE, 2021.

\bibitem{patterson2008effect}
M.~K.~Patterson,
``The effect of data center temperature on energy efficiency,''
in \emph{Proc. 11th Intersoc. Conf. Therm. Thermomech. Phenom.
Electron. Syst.}, Orlando, FL, USA, 2008, pp.~1167--1174.

\bibitem{Nazari13}
M. H. Nazari, M. Ilic, J. P. Lopes,
``Small-signal stability and decentralized control design for electric energy systems with a large penetration of distributed generators,"
\emph{Control Engineering Practice},
Volume 20, Issue 9,
2012,
Pages 823-831.

\bibitem{ecocenter2026}
A. Jahanshahi, S. R. Golrouye, O. Anderson, N. Yu, and D. Wong,
``Coordinating Power Grid Frequency Regulation Service with Data Center Load Flexibility,"
\emph{arXiv preprint arXiv:2601.22487}, Jan. 2026.

\end{thebibliography}
\end{document}